\documentclass[sigconf]{acmart}

\usepackage{booktabs} % For formal tables
\usepackage{upgreek} % For uppercase greek
\usepackage{tikz-cd}
\usepackage{tikz}
\usetikzlibrary{shapes,arrows}
\usepackage{hyperref}
\usepackage{graphicx,verbatimbox}
\usepackage{listings}
\usepackage{subcaption}
\usepackage{mathtools}
\usepackage{color}
\usepackage{tabularx,ragged2e}
\usepackage{multirow}
\usepackage{mathtools}
\usepackage{enumitem}
\usepackage[british]{babel}
\usepackage{xspace} 
\usepackage{placeins}
\usepackage{balance}

\copyrightyear{2022} 
\acmYear{2022} 
\setcopyright{licensedothergov}\acmConference[ADCS '22]{Australasian Document Computing Symposium}{December 15--16, 2022}{Adelaide, SA, Australia}
\acmBooktitle{Australasian Document Computing Symposium (ADCS '22), December 15--16, 2022, Adelaide, SA, Australia}
\acmPrice{15.00}
\acmDOI{10.1145/3572960.3572980}
\acmISBN{979-8-4007-0021-7/22/12}

\begin{document}

\title{Neural Rankers for Effective Screening Prioritisation in Medical Systematic Review Literature Search}
\author{Shuai Wang}
\affiliation{%
	\institution{University of Queensland}
	\city{Brisbane}
	\country{Australia}
}
\email{shuai.wang2@uq.edu.au}

\author{Harrisen Scells}
\affiliation{%
  \institution{Leipzig University}
  \city{Leipzig}
  \country{Germany}
}
\email{harry.scells@uni-leipzig.de}

\author{Bevan Koopman}
\affiliation{%
	\institution{CSIRO}
	\city{Brisbane}
	\country{Australia}
}
\email{b.koopman@csiro.com}

\author{Guido Zuccon}
\affiliation{%
  \institution{University of Queensland}
  \city{Brisbane}
  \country{Australia}
}
\email{g.zuccon@uq.edu.au}

\begin{abstract}

Medical systematic reviews typically require assessing all the documents retrieved by a search. The reason is two-fold: the task aims for ``total recall''; and documents retrieved using Boolean search are an unordered set, and thus it is unclear how an assessor could examine only a subset.
 %One consequence of this assessment process is that it massively increases the amount of time and money required to complete the systematic review. Recently, ranking the set of retrieved documents, or \textit{screening prioritisation}, has gained traction in the systematic review community. Presenting researchers with the most relevant documents can enable them to begin the downstream processes of the systematic review creation process earlier, and thus lead to an earlier completion of the review compared to when not using screening prioritisation, or even to stop assessing documents early.
\textit{Screening prioritisation} is the process of ranking the (unordered) set of retrieved documents, allowing assessors to begin the downstream processes of the systematic review creation earlier, leading to earlier completion of the review, or even avoiding screening documents ranked least relevant.

%Pre-trained language models have achieved state-of-the-art ranking effectiveness, outperforming traditional methods on many IR tasks such as passage ranking, question-answering and domain-specific search. They are yet to be shown effective for screening prioritisation.

Screening prioritisation requires highly effective ranking methods. Pre-trained language models are state-of-the-art on many IR tasks but have yet to be applied to systematic review screening prioritisation. In this paper, we apply several pre-trained language models to the systematic review document ranking task, both directly and fine-tuned. An empirical analysis compares how effective neural methods compare to traditional methods for this task. We also investigate different types of document representations for neural methods and their impact on ranking performance. 

Our results show that BERT-based rankers outperform the current state-of-the-art screening prioritisation methods. However, BERT rankers and existing methods can actually be complementary, and thus, further improvements may be achieved if used in conjunction.
%We also find that these neural rankers obtain high effectiveness in systematic review topics \todo{different than the state-of-the-art active learning strategies [unclear]}, meaning that it has the potential to achieve even higher performance when relevance feedback in active learning is also used for screening prioritisation.
	
\end{abstract}


\maketitle

\section{Introduction}
In medicine, systematic reviews are considered the most comprehensive and reliable instrument to synthesise evidence for a specific research question. When searching for documents for a systematic review, all retrieved documents are assessed to ensure the systematic review is comprehensive and correct. Arguably, however, the only reason that all documents must be assessed is that the documents retrieved from typical databases such as PubMed are returned as an unordered set, thus all equally relevant to the query.
A typical systematic review requires upwards and beyond 10,000 documents to be assessed~\cite{shemilt2016use}. Naturally, this assessment process, called \textit{screening}, is costly and is further exasperated given the requirement that the documents should be screened multiple times by different assessors to account for biases and disagreements~\cite{shemilt2016use, higgins2019cochrane}. Screening is executed by assessing the title and abstract of a retrieved publication. 
Once a document is considered relevant at a title-abstract level, it must also be assessed at the full-text level. Overall, it is a laborious process.

%To help systematic review researchers to achieve a more effective document screening, several research works have investigated algorithms to rank the documents obtained from search \cite{bibid}. The task that helps to rank documents in document screening is called screening prioritisation. Screening prioritisation can bring benefits in a systematic review document screening in two ways, including: (1) reviewers may exit early in the document screening when enough evidence has been found; (2) Parallel screening on abstract level and full-text level may bring further benefit as early identification of relevant articles in abstract level may allow other assessors to screen on full-text of the articles earlier. 

As a way to reduce systematic review creation effort, time delays and costs, the systematic review community has looked at adopting automation tools~\cite{o2015using}. One of the tasks for which automation tools can be helpful is \textit{screening prioritisation}.
%One set of methods that have seen some adoption by the systematic review community tackle the problem of \textit{screening prioritisation}. 
Here, retrieved documents are ordered by their relevance to the review. Screening prioritisation can reduce the time and cost factors associated with systematic review creation in two ways: (1) researchers can begin and complete downstream tasks, such as full-text screening in parallel, earlier than if publications were assessed in random order; and (2) researchers can stop screening early by only considering the top-$k$ ranked documents, possibly with a certain level of confidence that ``total recall'' (or some approximation of) has been achieved.
Numerous methods have been proposed by the Information Retrieval community to address this problem~\cite{lee2018seed, wang2022seed, cormack2017technology, grossman2011technology, 10.1145/3477495.3531748, cormack2019systems, grossman2017automatic, yang2022goldilocks, alharbi2018retrieving, alharbi2017ranking, scells2017qut, wu2018ecnu}. These methods can be separated into two classes: 
(1) methods that directly use queries to rank, where the queries can be the title~\cite{alharbi2017ranking, alharbi2018retrieving}, the review's Boolean query~\cite{alharbi2018retrieving, alharbi2017ranking, alharbi2019ranking, scells2020clf, wu2018ecnu}, the objectives of the review~\cite{scells2017qut, scells2017integrating}, or a set of studies known a priori~\cite{wang2022seed, lee2018seed, 10.1145/3477495.3531748}; and 
(2) use different ranking techniques, including relevance feedback~\cite{alharbi2019ranking}, and various forms of active learning~\cite{cormack2017technology, grossman2011technology, cormack2019systems, grossman2017automatic}, without the need for a query. 
While initial attempts to use new state-of-the-art pre-trained language models have been made in the second-ranking category of methods \cite{yang2022goldilocks}, no research has considered using pre-trained language models for the first, query-focused category; this is the focus of this paper. 
%As of yet, there has been no research in investigating new state-of-the-art pre-trained-language models for the first class of screening prioritisation, which is the focus of this paper.
\vspace{5pt}

In this paper, we investigate the use of rankers based on pre-trained-language models for the systematic review screening prioritisation task. Specifically, we apply different types of zero-shot pre-trained models to investigate the effectiveness of different pre-training types. We also fine-tune the ranker for the screening prioritisation task and use different fine-tuning approaches to investigate the effectiveness of fine-tuning. Lastly, we perform extensive analysis to show how the effectiveness of neural methods differs from current state-of-the-art non-neural screening prioritisation methods.
%
%In this work, we have summarised the following contribution:
%Through this paper, we have performed initial experiments and analysis to help researchers in the systematic review screening prioritisation task to understand the potential of using Neural models and how to use these models in the task. 
%
We make the following contributions: 

\vspace{5pt}

\begin{enumerate}
	\item We conduct experiments using multiple neural rankers both under zero-shot and fine-tuned settings. We compare these rankers with the state-of-the-art methods for the task to show that they significantly outperform the current best ranking methods. We obtain similar results to methods that use interactive ranking approaches with relevance feedback (i.e., active learning).

\vspace{5pt}

	\item We use two representations for candidate documents to fine-tune our pre-trained language models: \text{Title} (title only) and \text{TiAb} (title+abstract), and we show the effectiveness difference between these two representations. We find that encoding the abstract with the title (TiAb) is far better than title only to represent candidate documents.  %to see if encoding passage helps with ranking.

\vspace{5pt}

	\item We perform a query-by-query comparative analysis on the best neural and state-of-the-art methods. We show that although the average performance of fine-tuned BioBERT and the current state-of-the-art iterative methods obtain similar evaluation results, the effectiveness differs significantly at the topic level. The finding indicates that neural methods may achieve even higher performance if the current state-of-the-art active learning methods are combined with neural rankers.  
\end{enumerate}

%\vspace{-8pt}
\vspace{5pt}
\section{Related Work}

%\vspace{-7pt}
\noindent
\subsection{IR for Medical Systematic Review Automation}%
\vspace{5pt}
Medical systematic reviews follow a standardised process to ensure consistency and quality. The most laborious part of the process is the screening of documents~\cite{shemilt2016use}. These documents are retrieved using a complex Boolean query that attempts to precisely encode the information need required to answer the research question of the systematic review and to ensure the reproducibility of the review (i.e., re-running the search in the future should produce the same set of documents)~\cite{garritty2021cochrane, macfarlane2022search, scells2017collection}. The Boolean query retrieves an unordered set of documents. However, ranking the set of retrieved documents has two main advantages:
\begin{enumerate}[leftmargin=15pt,noitemsep,topsep=5pt]
	
\item Systematic review creation often comes with a budget, which means screening of all retrieved documents may be impossible with a small budget or limited time (i.e. rapid reviews)~\cite{marshall2019rapid, MICHELSON2019100443}. To help systematic reviewers obtain the most relevant documents within a budget, an effective ranking of documents can ensure that the most relevant documents are found without having to screen all documents.
\vspace{5pt}

\item Systematic reviews require a two-stage screening process: first, documents are assessed by only title and abstract; only if relevant are full-text documents reviewed.
These two steps are run sequentially, and for each document, full-text screening can start as soon as that document has undergone the title-abstract level screening. Therefore, exhausting the screening of all relevant documents before starting the screening of non-relevant ones (a task achieved through high-quality screening prioritisation) leads to downstream steps taking place in parallel with the remaining screening. This ultimately allows the whole systematic review to conclude earlier than if screening prioritisation was not implemented.

%Systematic review document screening often comes with a time limit \cite{bibid}. Therefore, it is typically achieved in parallel by multiple assessors \cite{bibid}. One way assessors work together is to assign one group of assessor to judge documents on abstract-level, while another group of assessor judge abstract-level relevant documents on full-text level \cite{bibid}. However, full-text document assessors can not start judging on documents until the first relevant documents been judged in abstract-level, and systematic review screening can not be finished until the last full-text documents are achieved.
%By implementing effective ranking of candidate documents, the initial wait time of full-text assessing can be minimized, so that time required for conducting the whole systematic review may be shortened. 
\end{enumerate}
\vspace{5pt}

To achieve effective ranking in systematic review document screening, research has investigated using (1) \textit{One-off Ranking:} methods that use queries to obtain one-off rankings of candidate documents~\cite{alharbi2017ranking, alharbi2018retrieving, scells2017qut, scells2017integrating, wang2022seed, lee2018seed, 10.1145/3477495.3531748}, and (2) \textit{Iterative Ranking:} methods that iteratively acquire user feedback during document screening and re-rank the unjudged documents \cite{cormack2017technology, grossman2011technology, cormack2019systems, grossman2017automatic}. These methods could be used in conjunction; e.g., start off with a one-off ranking and then update the ranking as feedback is received.

\vspace{5pt}

Aside from ranking, IR research has considered techniques for query improvement and reformulation that actually reduce the number of documents needing screening~\cite{wang2021mesh, wang2022automated, scells2021comparison, scells2020automatic, scells2019automatic}. Although these methods affect screening, and could be used in conjunction with screening prioritisation, they take place before the screening task, and thus we do not consider them further in this paper.

\subsection{Pre-trained Language Models}%
\vspace{5pt}

Advances in pre-trained language models like BERT~\cite{devlin2018bert}, Rob\-erta~\cite{liu2019roberta} and T5~\cite{raffel2020exploring} show effective on multiple downstream tasks: document ranking~\cite{gao2021condenser, wang2021bert, hang2022interpolate}, question-answering~\cite{qu2019bert}, and conversational search~\cite{qu2019bert, ferreira2022open}. These language models are pre-trained on large text corpora (Wikipedia~\cite{devlin2018bert} or PubMed~\cite{lee2020biobert, gu2021domain, peng2019transfer}) to learn textual features such as sentence structure, word semantics and sentence semantics, before being applied to specific tasks, such as document ranking. There are typically two ways to apply pre-trained language models to downstream tasks: (1) apply the model directly to the task, i.e., \textit{zero-shot}; or (2) using training samples from the downstream task to \textit{fine-tune} the pre-trained language model.

\vspace{5pt}

Currently, these neural rankers have yet to be investigated for the systematic review screening prioritisation task. The only related work is \citet{yang2022goldilocks}, which trains classifiers based on pre-trained language models to perform automatic Technology Assisted Reviews in domains other than systematic review literature search. In this paper, we propose methods and deliver initial experiments to examine both zero-shot and fine-tuned neural methods for the screening prioritisation task in the systematic review and their corresponding effectiveness, thus filling this gap.

\section{Neural Rankers for Screening Prioritisation}

\subsection{Model Architecture}
%\todo{@harry: this section would be better structured to first introduce the two model architectures and then have two subsubsections that describe what is done for both, and then another subsubsection for representations}

In this paper, we examine two different avenues for using pre-trained language model based rankers for the screening prioritisation task: (1) zero-shot and (2) fine-tuned. Both approach rely on the typical monoBERT cross-encoder architecture to compute scores~\cite{nogueira2019multi}. For each query-document pair, we concatenate the text of a document $d$ with that of the query $q$ (separated by a $SEP$ token) and encode this input to obtain the relevance score of $d$ given $q$.

In the \textit{zero-shot} setting, we use the pre-trained language models directly on the screening prioritisation task. In our experiments, we consider the BERT base model~\cite{devlin2018bert}, the BERT base model fine-tuned on the MS MARCO dataset~\cite{gao2021rethink}, and an array of BERT models pre-trained on different medical-specific text corpora: BioBERT~\cite{lee2020biobert}, PubMedBERT~\cite{gu2021domain} and BlueBERT~\cite{peng2019transfer}.

In the \textit{fine-tuned} setting, we further fine-tune the BERT model using the training portion in our dataset and then apply the resulting ranker on the screening prioritisation task on the test portion of the dataset. For fine-tuning, we experiment with 
BERT base, BERT fine-tuned on MS MARCO and BioBERT. We use localised contrastive loss with triples of $<\text{title}, D^+, set(D^-) >$ where $D$ is the representation of a candidate document, $D^+$ is a document judged relevant at the abstract level, $D^-$ is conversely a document judged not-relevant at the abstract level. We use the framework proposed by Gao et al.~\cite{gao2021rethink} and default learning parameters to develop our model.

\subsection{Documents Representation}
\label{doc_rep}
We also investigate two ways to represent a document: (1) \textit{Title}, and (2) Title and Abstract (\textit{TiAb}) concatenated and separated by BERT's $[SEP]$ token. 
Using the \textit{Title} representation instead of the \textit{TiAb} may be reasonable because BERT has an input limit of 512 tokens; concatenating the title and abstract with the query may exceed the length of the BERT input, and the text input will then be truncated (i.e., the exceeded tokens discarded). This truncation may remove information in some of the long training samples, and possibly cause training mismatch as some samples may be truncated and others not. A shorter input size also results in slightly faster inference (i.e., reduced latency).  
We perform the comparison between the two representations for the fine-tuned setting only; for the zero-shot setting, we only investigate the \textit{TiAb} representation -- this is because our early experiments showed this to be highly ineffective for zero-shot; thus, we do not retain this of interest for the paper.
%We use the \textit{Title} representation only on fine-tuned Neural models to compare their effectiveness difference. 
%There are two intuitions for why using the \textit{Title} representation instead of the \textit{Tiab} may be reasonable: 

%\begin{enumerate}
%	\item BERT has an input limit of 512 tokens; concatenating the title and abstract with the query may exceed the length of the BERT input, and the text input will then be truncated (i.e., the exceeded tokens discarded). This truncation may remove information in some of the long training samples, and possibly cause training mismatch as some samples may be truncated and others not.
%	
%	\item Length of BERT input tokens can be reduced using only the title of the  document. A shorter input size can result in faster computation and ranking of  documents.
%\end{enumerate}

\section{Experimental Setup}

\subsection{Dataset \& Evaluation}
\label{dataset}
We use three CLEF Technological Assisted Review (TAR) datasets to evaluate the effectiveness of the screening prioritisation task~\cite{kanoulas2017clef, kanoulas2018clef, kanoulas2019clef}.
The CLEF TAR datasets contain 50 systematic review topics in 2017 (20 training, 30 testing)~\cite{kanoulas2017clef} and 80 topics in 2018 (50 training, 30 testing)~\cite{kanoulas2018clef}. All topics in CLEF TAR 2017 and 2018 datasets are Diagnostic Test Accuracy (DTA) systematic reviews. The CLEF TAR 2019 dataset includes 40 intervention systematic review topics (20 training, 20 testing), 88 DTA topics (80 training, 8 testing), one prognosis review and one qualitative review~\cite{kanoulas2019clef}. 
For each topic, the collection contains the title of the review, the Boolean query used for document retrieval, the documents (retrieved by issuing the Boolean query), and the relevance labels for the documents (abstract-level and full-text level).
Our experiments use the training and testing portions pre-defined in the above datasets. We use all systematic review topics from CLEF TAR 2017 and 2018. For CLEF TAR 2019, we only use intervention (2019-intervention) and DTA (2019-dta) topics as there is no training data for the other types of reviews.

We use the title of the review for each topic as the query to rank documents. In the CLEF TAR dataset, documents are represented by their PubMed PMID (i.e., a document identifier). In our experiment, we obtain the title and abstract of these documents directly from the PubMed index.

We consider only abstract-level relevance for our evaluation, as PubMed full-text is not publicly available from its index. We use the same evaluation measures as CLEF TAR: the rank position of the last relevant document in the ranking (Last\_Rel), AP, Recall@$p$\% ($p = {1, 5, 10, 20}$), and Work Saved Over Sampling (WSS) at $k$\% ($k = {95\%, 100\%}$).  %The intuition of WSS assumes that removing documents in the lower ranks still ensures that $k$ percentage documents are screened. 
Intuitively, WSS@$k$ measures the fraction of the screening workload saved if one stops examining the ranking once $k\%$ of the relevant documents have been found compared to screening the entire set of documents \cite{cohen2006reducing}. 
%The function to calculate work saved by sampling is using $\frac{|C|-\text{Rel@k}}{|C|}$.
%Where $C$ is the number of studies originally retrieved (i.e., the  set for re-ranking), 
%$Rel@k$ is the rank position of the document when k\% of relevant documents have been retrieved. 
%Note that we use the same evaluation measures that have also been used in CLEF Technological-Assisted-Review task 2 in 2018 and 2019. 
All evaluation measures are computed using the script provided in CLEF TAR 2018.

\subsection{Baseline Methods}
We use traditional exact word-matching methods, including the Query Likelihood Model (QLM) and BM25, as baseline methods. To compute the QLM score of a query-document pair, we use Jelinek-Mercer (JM) smoothing~\cite{zhai2016text}. To compute BM25 scores, we used the Gensim toolkit~\cite{vrehuuvrek2011gensim}.

Additionally, for each CLEF TAR dataset, we obtained the best-performing runs from the campaign results and available data --- we deem these as being the current state-of-the-art (noting there does not seem to be follow-up work that significantly outperformed these runs). Note that we can only use runs that do not cut off the document list for a fair comparison. The runs selected are then:
\begin{itemize}
	\item CLEF TAR 2017: \path{waterloo.A-rank-normal} \cite{cormack2017technology};
	\item CLEF TAR 2018: \path{UWB} \cite{cormack2018technology};
	\item CLEF TAR 2019 dta:  \path{Sheffield/DTA/DTA_sheffield-Odds_Ratio} \cite{alharbi2019ranking};
	\item CLEF TAR 2019 Intervention: \path{Sheffield/DTA/DTA_sheffield-Log_Likellihood} \cite{alharbi2017ranking}.
\end{itemize}

CLEF TASK participants were allowed to use \textit{Iterative Ranking}, that is, use the relevance assessments of the topics for explicit relevance feedback. Thus simulating the user provides feedback on results in an interactive manner.  %and were evaluated using the same metrics. The feedback was obtained by providing true relevance judgement by the task organizers. 
However, we only consider the setting of the screening prioritisation task with no feedback. For the CLEF TAR 2017 and 2018 datasets, we could identify the runs that did not use feedback via the task overview papers. The selected runs for CLEF-2017 and CLEF-2018 are:
\begin{itemize}
	\item CLEF TAR 2017: \path{sheffield.run4} \cite{alharbi2017ranking};
	\item CLEF TAR 2018: \path{shef-general} \cite{alharbi2018retrieving};
\end{itemize}
However, for CLEF TAR 2019, we could not distinguish between feedback and non-feedback runs; therefore, we consider the overall best run as the state-of-the-art, noting this run may or may not be using feedback.

\subsection{Fine-tuning Details}

In our experiments that involve fine-tuning, we fine-tune the BERT models using a group size of 10: for each training step, one positive sample and nine negative samples are used to compute the loss. 
We use a batch size of 3. Once the models are fine-tuned, we concatenate the systematic review title with the representation of the document to compute their relevance scores. In the experiments, we report the models' results using the last model checkpoint that has been fine-tuned for 100 epochs. We also discuss model convergence by investigating the test effectiveness on every saved checkpoint (saved every 100 training steps).

\section{Results}

%We show results on four datasets, including CLEF-2017, CLEF-2018, CLEF-2019-dta and CLEF-2019-intervention.

\begin{table*}
	\centering
%	\small
	\begin{tabular}{ll|ll|llll|ll}
		\toprule
		%\multicolumn{2}{c|}{Dataset}&\multicolumn{7}{c}{2017}\\ \midrule
		\multicolumn{1}{l}{Dataset}&\multicolumn{1}{l|}{Method}&\multicolumn{1}{c}{Last\_Rel}&\multicolumn{1}{c|}{AP}&\multicolumn{1}{c}{Recall@1\%}&\multicolumn{1}{c}{Recall@5\%}&\multicolumn{1}{c}{Recall@10\%}&\multicolumn{1}{c|}{Recall@20\%}&\multicolumn{1}{c}{WSS95}&\multicolumn{1}{c}{WSS100} \\ \midrule
		
		\multirow{7}{*}{2017}&BM25&2999.7000&0.1497$^\dagger$&0.0931$^\dagger$&0.2717$^\dagger$&0.3851$^\dagger$&0.5737$^\dagger$&0.3518&0.2520\\
		&QLM&2999.5333&\textbf{0.1721}$^\dagger$&\textbf{0.1071}$^\dagger$&\textbf{0.2849}$^\dagger$&\textbf{0.4067}$^\dagger$&\textbf{0.6340}$^\dagger$&0.3588&0.2588\\ \cmidrule{2-10}
		&BERT&\textbf{2898.0333}&0.0917&0.0119&0.1052&0.2594&0.4386&0.3565&\textbf{0.3082}\\
		&BERT-M&3161.5667&0.1247&0.0220&0.1324&0.2682&0.5265&\textbf{0.3627}&0.2606\\
		&BioBERT&3358.2667&0.0998&0.0301&0.1801&0.2922&0.4372&0.2373$^\dagger$&0.1836$^\dagger$\\
		&BlueBERT&3824.4667$^\dagger$&0.0365$^\dagger$&0.0001$^\dagger$&0.0113$^\dagger$&0.0349$^\dagger$&0.0935$^\dagger$&0.0233$^\dagger$&0.0407$^\dagger$\\
		&PubMedBERT&3628.5000$^\dagger$&0.0670&0.0156&0.0667&0.1433$^\dagger$&0.2717$^\dagger$&0.1205$^\dagger$&0.1018$^\dagger$\\
		\midrule
		
		\multirow{7}{*}{2018}&BM25&6095.2000&\textbf{0.1683}$^\dagger$&\textbf{0.0972}$^\dagger$&\textbf{0.2947}$^\dagger$&0.4444$^\dagger$&0.6441$^\dagger$&0.4202&0.2336\\
		&QLM&5956.9667&0.1660$^\dagger$&0.0896$^\dagger$&0.2904$^\dagger$&\textbf{0.4544}$^\dagger$&\textbf{0.6456}$^\dagger$&0.4322&0.2540\\ \cmidrule{2-10}
		&BERT&6158.2333&0.1249&0.0222&0.1348&0.2960&0.5422&0.4037&\textbf{0.2622}\\
		&BERT-M&\textbf{5805.4667}&0.1398&0.0188&0.1465&0.2997&0.5894&\textbf{0.4476}&0.2567\\
		&BioBERT&6696.9333&0.0881&0.0236&0.1526&0.2580&0.4059$^\dagger$&0.1499$^\dagger$&0.0836$^\dagger$\\
		&BlueBERT&7204.3667$^\dagger$&0.0329$^\dagger$&0.0014$^\dagger$&0.0093$^\dagger$&0.0198$^\dagger$&0.0493$^\dagger$&0.0017$^\dagger$&0.0193$^\dagger$\\
		&PubMedBERT&6955.8333$^\dagger$&0.0700$^\dagger$&0.0112&0.0636$^\dagger$&0.1593$^\dagger$&0.3138$^\dagger$&0.1424$^\dagger$&0.0883$^\dagger$\\
		\midrule
		
		\multirow{7}{*}{2019-dta}&BM25&2722.7500&0.1185&0.0479&0.2129&\textbf{0.3290}&0.5276&0.3138&0.2080\\
		&QLM&\textbf{2318.2500}&\textbf{0.1223}&\textbf{0.0644}&\textbf{0.2164}&0.3270&\textbf{0.5335}&\textbf{0.3470}&\textbf{0.2477}\\ \cmidrule{2-10}
		&BERT&2513.8750&0.0922&0.0244&0.1318&0.2381&0.3906&0.2577&0.2095\\
		&BERT-M&3233.7500&0.0955&0.0105&0.0792&0.1979&0.3793&0.2629&0.1232\\
		&BioBERT&3264.0000&0.0810&0.0160&0.1290&0.2294&0.3365&0.1370&0.0950\\
		&BlueBERT&3771.0000&0.0688&0.0010&0.0256&0.0526&0.1050&0.0227&0.0160\\
		&PubMedBERT&3330.2500&0.1044&0.0335&0.1226&0.2144&0.3119&0.2016&0.0979\\
		\midrule
		
		\multirow{7}{*}{2019-int.}&BM25&1715.6000&0.2112&0.0968&\textbf{0.3053}&\textbf{0.3989}&\textbf{0.5542}&0.3510&0.2955\\
		&QLM&1724.0500&\textbf{0.2123}&\textbf{0.0981}&0.2793&0.3851&0.5110&0.3397&0.2939\\ \cmidrule{2-10}
		&BERT&\textbf{1398.5500}&0.1603&0.0536&0.2104&0.3282&0.5041&\textbf{0.3624}&\textbf{0.3330}\\
		&BERT-M&1836.2000&0.1769&0.0384&0.1951&0.3545&0.5268&0.3228&0.2663\\
		&BioBERT&1832.8500&0.1463&0.0530&0.1346&0.1982&0.3074$^\dagger$&0.1585$^\dagger$&0.1631$^\dagger$\\
		&BlueBERT&2057.0000&0.0462&0.0063&0.0275$^\dagger$&0.0513$^\dagger$&0.1066$^\dagger$&0.0083$^\dagger$&0.0361$^\dagger$\\
		&PubMedBERT&1974.2500&0.0780&0.0124&0.0502&0.0905$^\dagger$&0.2748$^\dagger$&0.1207$^\dagger$&0.0944$^\dagger$\\
		\bottomrule
	\end{tabular}
	\caption{Results obtained using pre-trained language models in a zero-shot setting. Statistical significant differences (Student's two-tailed paired t-test with Bonferonni correction, p < 0.05) between BERT and all other methods are indicated by $\dagger$.}
	\label{table:search_result_zero_shot}
\end{table*}

\subsection{Zero-shot Prioritisation}
Table \ref{table:search_result_zero_shot} reports the effectiveness of the neural methods under the zero-shot setting, along with that of the baselines considered. In the table, BERT-M refers to BERT fine-tuned on the MS MARCO dataset\footnote{We consider this a zero-shot method because fine-tuning is done for a different dataset than the target one, and a task of Adhoc web search is different to screening prioritisation.}. For these experiments, we consider the title and abstract (TiAb) representation for documents.
%We show zero-shot result of Neural methods in Table \ref{table:search_result_zero_shot}. First, we compare Neural methods with baseline methods. 

QLM and BM25 outperform the zero-shot neural methods across all evaluation metrics, with a few exceptions: (1) the zero-shot BERT outperforms the other methods in all datasets but 2019-dta for WSS100, in 2019-intervention for WSS95, and in 2017 and 2019-intervention for Last\_Rel;  (2) the zero-shot BERT-M outperforms the other methods in 2017 and 2018 for WSS95, and in 2018 for Last\_Rel.

Comparing the different zero-shot neural rankers, we find that generally, BERT and BERT-M perform similarly and better than the remaining domain-specific models; among these remaining models, BioBERT is the one that performs best. We further note that the fine-tuning on the MS MARCO dataset (BERT-M), and thus to a different but related task to screening prioritisation, does not lead to significant effectiveness boosts compared to BERT. Overall, the use of zero-shot neural rankers for the task of screening prioritisation does not appear to be a competitive and viable approach to the task.

% obtains the highest effectiveness in most evaluation metrics in our experiment, including Last\_Rel, Recall@20\%, WSS95 and WSS100. When the BERT base model is fine-tuned on MS Marco dataset (BERT-M), no significant effectiveness boost has been shown on any metrics. 
%
%Furthermore, models pre-trained on PubMed (BioBERT, BlueBERT, PubMedBERT) do not perform better than the BERT model. This finding indicates that Neural models are not necessary to be pre-trained on the same corpus to maintain a higher performance on zero-shot systematic review document prioritisation.

\subsection{Fine-tuned Prioritisation}
\label{sec_fine_tuned}

\begin{table*}[t!]
	\centering
%	\small
	\begin{tabular}{ll|ll|llll|ll}
		\toprule
		%\multicolumn{2}{c|}{Dataset}&\multicolumn{7}{c}{2017}\\ \midrule
		\multicolumn{1}{l}{Dataset}&\multicolumn{1}{l|}{Method}&\multicolumn{1}{c}{Last\_Rel}&\multicolumn{1}{c|}{AP}&\multicolumn{1}{c}{Recall@1\%}&\multicolumn{1}{c}{Recall@5\%}&\multicolumn{1}{c}{Recall@10\%}&\multicolumn{1}{c|}{Recall@20\%}&\multicolumn{1}{c}{WSS95}&\multicolumn{1}{c}{WSS100} \\ \midrule

		\multirow{7}{*}{2017}&BEST-No-Feedback&2382.4667&0.2179$^\dagger$&0.1308&0.3325$^\dagger$&0.4993$^\dagger$&0.6877$^\dagger$&0.4880$^\dagger$&0.3946$^\dagger$\\
		&BEST-Iterative&1469.4000&\textbf{0.3183}&0.1707&\textbf{0.5434}&\textbf{0.7322}&\textbf{0.8863}&\textbf{0.7009}&\textbf{0.6106}\\\cmidrule{2-10}
		&BM25&2999.7000$^\dagger$&0.1497$^\dagger$&0.0931$^\dagger$&0.2717$^\dagger$&0.3851$^\dagger$&0.5737$^\dagger$&0.3518$^\dagger$&0.2520$^\dagger$\\
		&QLM&2999.5333$^\dagger$&0.1721$^\dagger$&0.1071$^\dagger$&0.2849$^\dagger$&0.4067$^\dagger$&0.6340$^\dagger$&0.3588$^\dagger$&0.2588$^\dagger$\\\cmidrule{2-10}
		&BERT-Tuned&2418.8333$^\dagger$&0.2273$^\dagger$&0.1066$^\dagger$&0.4088$^\dagger$&0.5888$^\dagger$&0.7789&0.5435&0.4340$^\dagger$\\
		&BERT-M-Tuned&2475.2000$^\dagger$&0.2770&0.1419&0.3727$^\dagger$&0.5843$^\dagger$&0.7707&0.5234$^\dagger$&0.4187$^\dagger$\\
		&BioBERT-Tuned&\textbf{1461.6667}&0.3078&\textbf{0.1845}&0.4903&0.6816&0.8355&0.6530&0.5913\\

		\midrule

		\multirow{7}{*}{2018}&BEST-No-Feedback&5519.2000&0.2584$^\dagger$&0.1287$^\dagger$&0.3827$^\dagger$&0.5449$^\dagger$&0.7295$^\dagger$&0.5520$^\dagger$&0.4314$^\dagger$\\
		&BEST-Iterative&\textbf{2655.0000}&0.3776&0.1854&0.5940&\textbf{0.7696}&\textbf{0.9149}&\textbf{0.7558}&\textbf{0.6104}\\\cmidrule{2-10}
		&BM25&6095.2000$^\dagger$&0.1683$^\dagger$&0.0972$^\dagger$&0.2947$^\dagger$&0.4444$^\dagger$&0.6441$^\dagger$&0.4202$^\dagger$&0.2336$^\dagger$\\
		&QLM&5956.9667$^\dagger$&0.1660$^\dagger$&0.0896$^\dagger$&0.2904$^\dagger$&0.4544$^\dagger$&0.6456$^\dagger$&0.4322$^\dagger$&0.2540$^\dagger$\\\cmidrule{2-10}
		&BERT-Tuned&5581.7667&0.3467$^\dagger$&0.1981$^\dagger$&0.5028$^\dagger$&0.6772$^\dagger$&0.8196$^\dagger$&0.6188$^\dagger$&0.4121$^\dagger$\\
		&BERT-M-Tuned&5185.5000&0.3387$^\dagger$&0.1934$^\dagger$&0.4833$^\dagger$&0.6515$^\dagger$&0.8265$^\dagger$&0.6559&0.4815$^\dagger$\\
		&BioBERT-Tuned&4108.4000&\textbf{0.4444}&\textbf{0.2768}&\textbf{0.5975}&0.7574&0.8946&0.7194&0.6103\\\midrule
		
		\multirow{6}{*}{2019-dta}&BEST&2183.5000&0.2477&0.1685&0.4391&0.5940&0.7421&0.4899&0.3470\\ \cmidrule{2-10}
		&BM25&2722.7500&0.1185$^\dagger$&0.0479&0.2129&0.3290$^\dagger$&0.5276$^\dagger$&0.3138$^\dagger$&0.2080$^\dagger$\\
		&QLM&2318.2500&0.1223$^\dagger$&0.0644&0.2164&0.3270$^\dagger$&0.5335$^\dagger$&0.3470$^\dagger$&0.2477$^\dagger$\\ \cmidrule{2-10}
		&BERT-Tuned&1399.3750&0.2234&0.1580&0.4390&0.6013&0.7620&0.5870&0.4600\\
		&BERT-M-Tuned&1178.0000&0.2535&0.2049&0.4474&0.5904&0.7536&0.6151&0.4997\\
		&BioBERT-Tuned&\textbf{852.7500}&\textbf{0.3177}&\textbf{0.2604}&\textbf{0.4998}&\textbf{0.6710}&\textbf{0.8171}&\textbf{0.6857}&\textbf{0.5845}\\
		
		\midrule
		
		\multirow{6}{*}{2019-int.}&BEST&1132.0000&0.2929$^\dagger$&0.1655&0.4192$^\dagger$&0.5424$^\dagger$&0.7225$^\dagger$&0.4582$^\dagger$&0.3808$^\dagger$\\ \cmidrule{2-10}
		&BM25&1715.6000&0.2112$^\dagger$&0.0968$^\dagger$&0.3053$^\dagger$&0.3989$^\dagger$&0.5542$^\dagger$&0.3510$^\dagger$&0.2955$^\dagger$\\
		&QLM&1724.0500&0.2123$^\dagger$&0.0981$^\dagger$&0.2793$^\dagger$&0.3851$^\dagger$&0.5110$^\dagger$&0.3397$^\dagger$&0.2939$^\dagger$\\ \cmidrule{2-10}
		&BERT-Tuned&1374.3000&0.2808$^\dagger$&0.1646$^\dagger$&0.3736$^\dagger$&0.5274$^\dagger$&0.6586$^\dagger$&0.3629$^\dagger$&0.3011$^\dagger$\\
		&BERT-M-Tuned&1571.2500&0.3343$^\dagger$&0.1614&0.4021$^\dagger$&0.5649$^\dagger$&0.7061$^\dagger$&0.4461$^\dagger$&0.3623$^\dagger$\\
		&BioBERT-Tuned&\textbf{706.8500}&\textbf{0.4559}&\textbf{0.2155}&\textbf{0.5805}&\textbf{0.7374}&\textbf{0.8417}&\textbf{0.6462}&\textbf{0.5794}\\
		
		\bottomrule
	\end{tabular}
\vspace{6pt}

	\caption{Results obtained when using pre-trained language models in the fine-tuned setting. Statistical significant differences (Student's two-tailed paired t-test with Bonferonni correction, p < 0.05) between BioBERT-Tuned and all other methods are indicated by $\dagger$.}
	\label{table:search_result_tiab}
\end{table*}

%We further analyse evaluation results on Fine-tuned BioBERT, BERT and BERT-M models in Table~\ref{table:search_result_tiab}. 

Table~\ref{table:search_result_tiab} reports the effectiveness of the neural rankers under the fine-tuned settings. For these experiments, we only considered fine-tuning BERT, BERT-M and BioBERT, the three best-performing rankers in the zero-shot setting. The results in the table refer to using the title and abstract (TiAb) representation.

Firstly, we observe that the fine-tuning regime greatly improves the effectiveness of the considered neural rankers over the zero-shot setting of Table~\ref{table:search_result_zero_shot}. While this may be somewhat expected, we highlight that the considered datasets contain only a small portion of training samples available for fine-tuning --- thus, even this small signal is enough to train far more effective neural rankers.

% that model effectiveness has been significantly improved over fine-tuning, shown as performance increase on all evaluation metrics compared with the corresponding zero-shot Neural methods. Reminding that we only contain a small portion of training samples, which we showed in section \ref{dataset}. This finding indicates that Neural models should be fine-tuned in systematic review screening prioritisation tasks even though the training sample is typically small to achieve a significant effectiveness boost.

Next, we compare the three pre-trained language rankers. Even though BERT and BERT-M obtained higher effectiveness in the zero-shot setting, when fine-tuning is performed, BioBERT often obtains higher effectiveness among the three rankers.

%Next, by comparing three fine-tuned pre-trained language models, we find that even though the BERT models obtain higher effectiveness in the zero-shot setting, BioBERT often obtains higher effectiveness when the model is fine-tuned. This finding provides insights that pre-trained Neural models under the target domain may benefit model performance only if model fine-tuning is conducted.

We now turn our attention to comparing the neural rankers with the other considered baselines. The fine-tuned neural rankers are now able to outperform the baseline methods across all measures and datasets significantly. This finding strengthens our previous insights that pre-trained neural models should be fine-tuned for the systematic review screening prioritisation task, even if just on a few training samples.

%By comparing fine-tuned Neural models with baseline methods BM25 and QLM, we find that fine-tuned Neural methods outperform baseline methods significantly. This contradicts our finding in zero-shot Neural model results, where we find that the baseline methods generally outperform zero-shot Neural models. This finding strengthens our previous insights that pre-trained Neural models should be fine-tuned even on a few training samples to increase the effectiveness in the systematic review screening prioritisation task.

Table~\ref{table:search_result_tiab} also reports the effectiveness of the best no-feedback methods submitted to the respective CLEF TAR tasks\footnote{Recall that it is not possible to determine a CLEF 2019 run uses feedback data or not.}. We find that all the fine-tuned neural rankers considered here achieve higher effectiveness than the best no-feedback runs for the corresponding CLEF year, with minor exceptions. 
The table also reports the effectiveness of the best iterative run (i.e., feedback via relevance assessment) at CLEF 2017 and 2018. We find that the fine-tuned BioBERT ranker consistently achieves higher effectiveness than the best iterative runs for Recall@1\%, Recall@5\%, AP and WSS100. 
We also find that the difference in effectiveness is significantly higher for shallow evaluation metrics while it is marginally higher for deep evaluation metrics. 
This finding can be explained by the fact that the iterative runs use feedback and some form of active learning, and thus, while effectiveness may be low in the first rank positions\footnote{Which correspond to the first few iterations.}, as shown for example by Recall@1\% and 5\%, they improve as more feedback is accumulated.
This point is interesting because it suggests that the neural rankers investigated here, which are highly effective, especially at the top of the ranking, could be used within an iterative, active learning loop in a bid to further improve effectiveness throughout the whole of the ranking.

\subsection{Model Convergence}
\label{sec:conv}

\begin{figure*}
	%	\begin{minipage}{\textwidth}
		\centering
		\includegraphics[width=0.49\columnwidth]{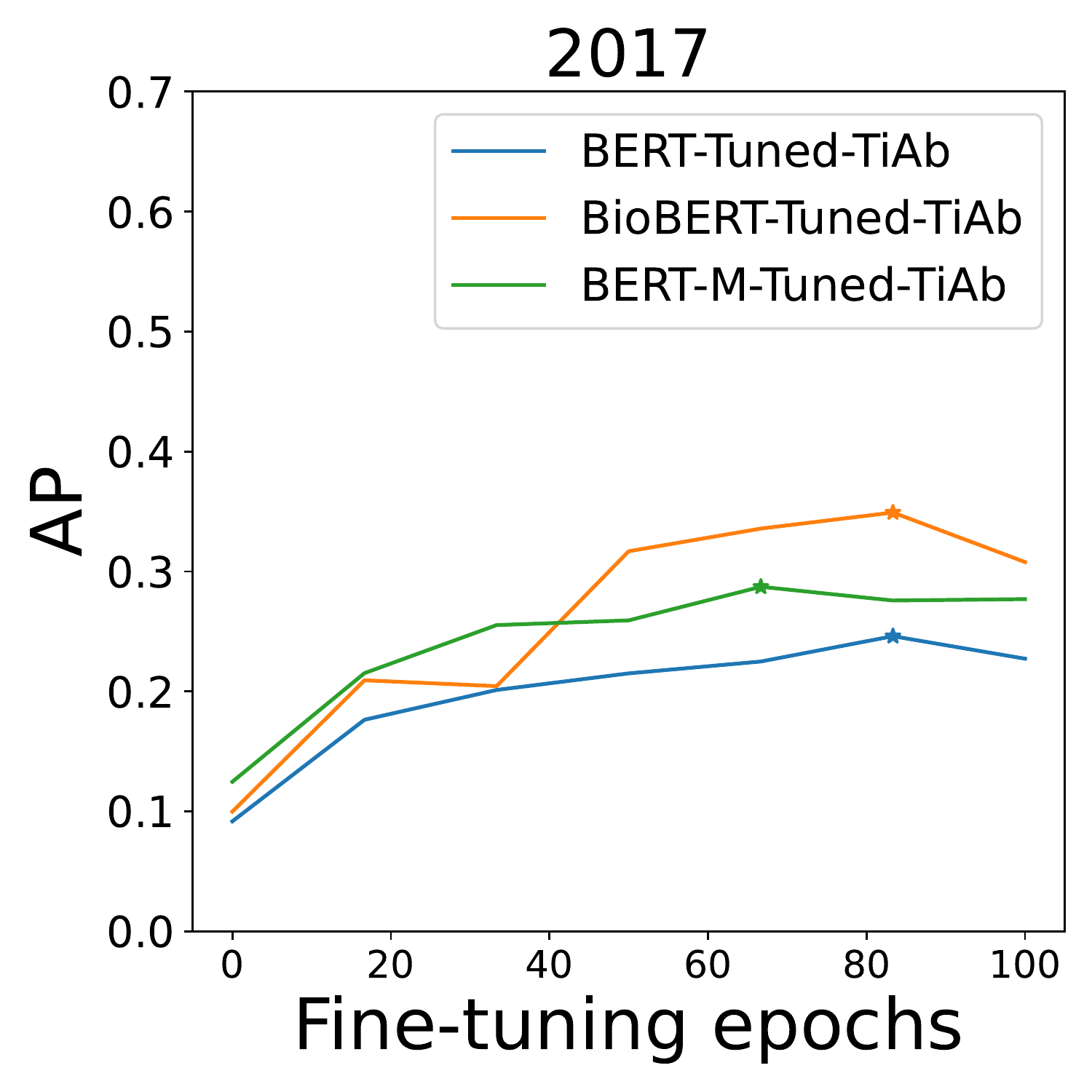}
		\includegraphics[width=0.49\columnwidth]{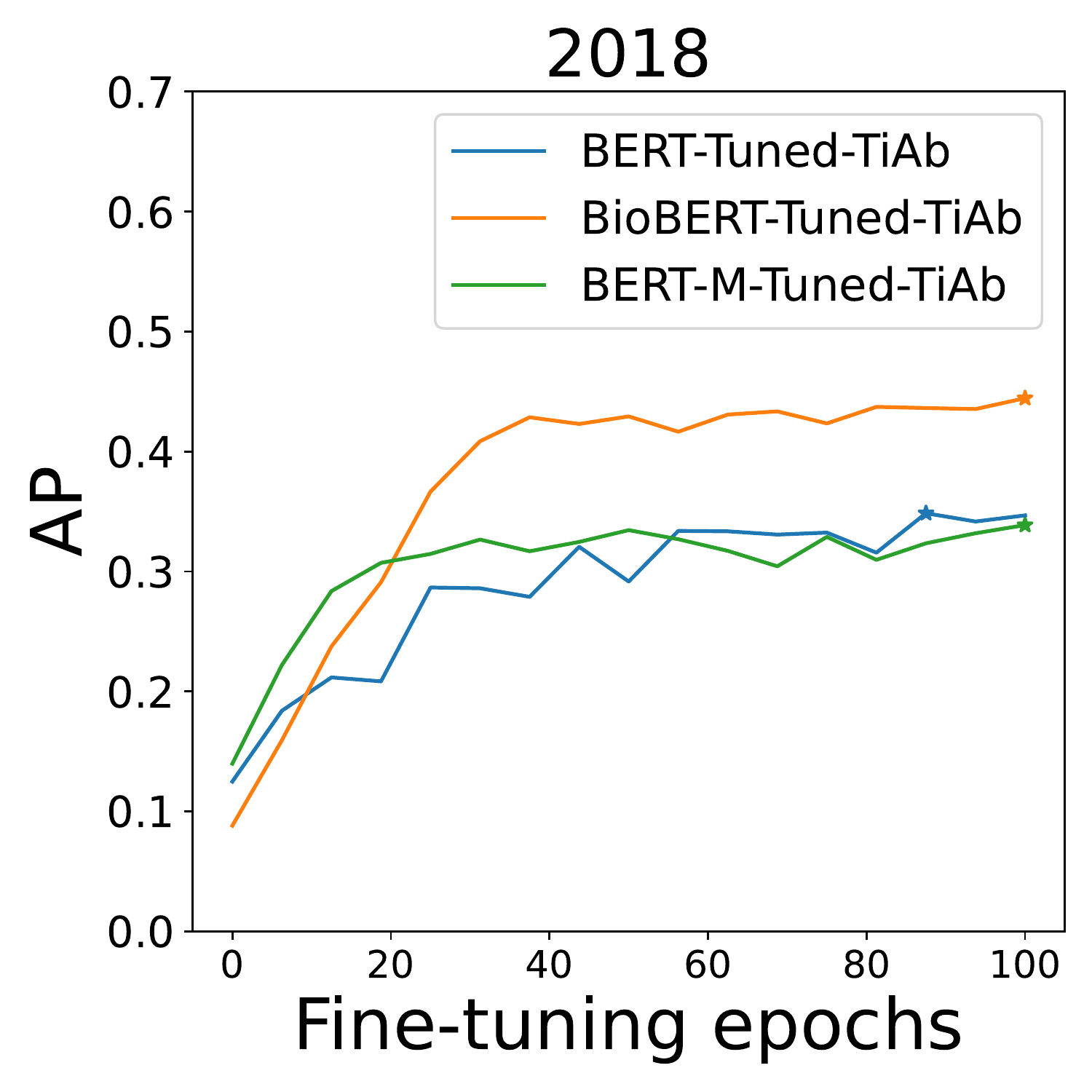}
		\includegraphics[width=0.49\columnwidth]{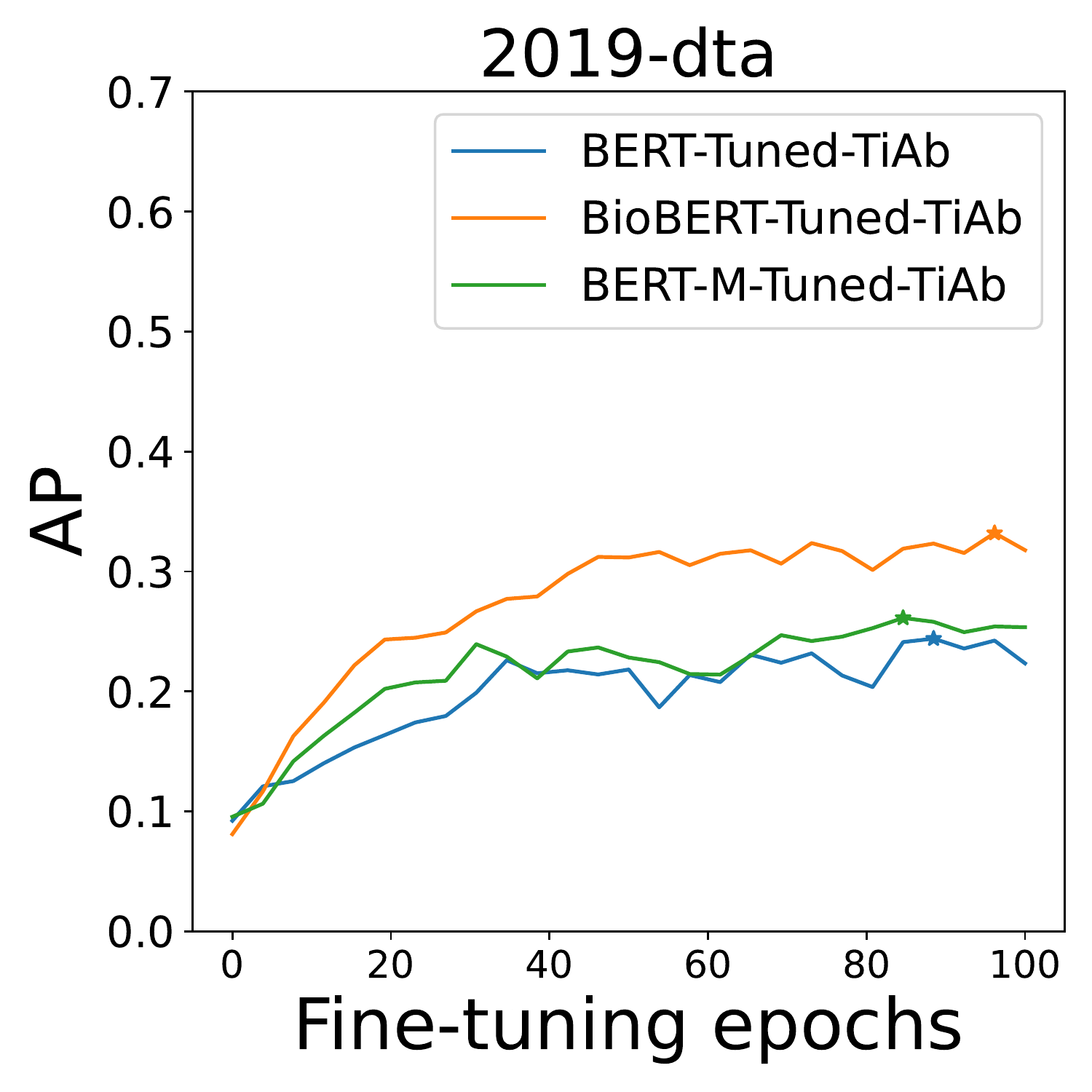}
		\includegraphics[width=0.49\columnwidth]{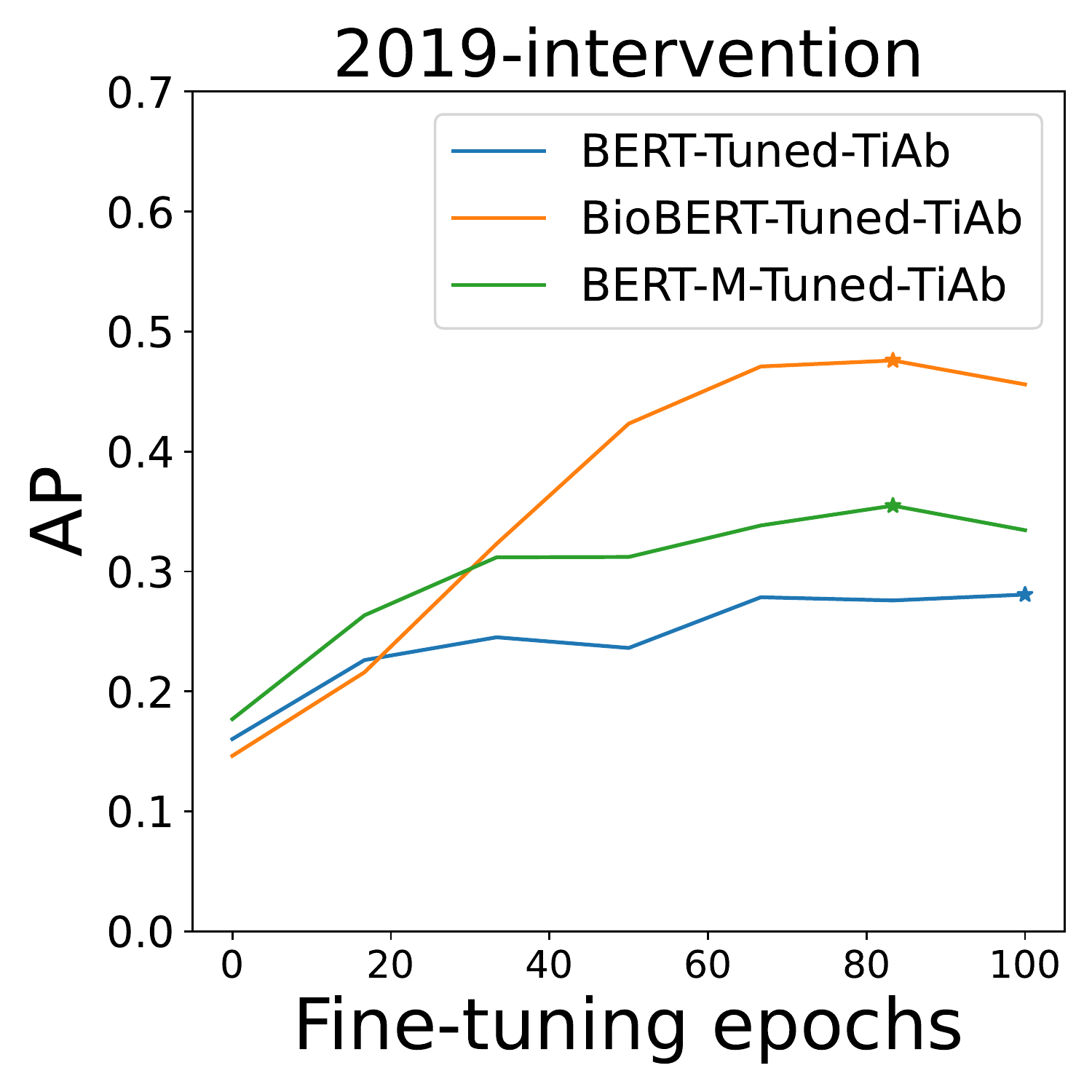}
		%	\end{minipage}
	
	\caption{Convergence of neural rankers during fine tuning. The y-axis reports AP measured on the test set, while the x-axis corresponds to subsequent fine-tuning steps. AP measurements are taken every 100 training steps. For each neural ranker, the checkpoint with the highest test AP is marked with $*$. }
	\label{fig:convergence}
	
\end{figure*}

We investigate the neural rankers' convergence during the fine-tuning process. Figure~\ref{fig:convergence} plots the AP values achieved on the test sets across subsequent steps of fine-tuning. From the figure, we deduce that ranker fine-tuning effectiveness saturates after about 100 epochs; this is confirmed by a statistical analysis that shows no significant differences are found in terms of test effectiveness across rankers' checkpoints beyond this training step (paired two-tailed t-test with Bonferroni correction, $p<0.05$).

We further note that the results reported in Table~\ref{table:search_result_tiab} are not necessarily the best results these neural rankers could achieve. In fact, in Table~\ref{table:search_result_tiab}, for each ranker, we reported the effectiveness of the last checkpoint	-- obtained after 100 fine-tuning epochs. However, the best test effectiveness is actually achieved by earlier checkpoints. If an effective way to detect the optimal point at which fine-tuning should be stopped, higher test effectiveness than that reported in Table~\ref{table:search_result_tiab} would be achieved. We note that one approach to this is using a validation dataset (though not guaranteeing optimal convergence): checkpoints would be measured against this dataset, and fine-tuning stopped after improvements below a threshold have been observed. No validation set was provided in CLEF, and we decided that splitting the training set to have a small validation set would have resulted in too little data for training (and too little data for validation, making validation unreliable).

%	In addition, the result shows that theoretically, the fine-tuned neural models still have the potential to achieve even higher effectiveness, as the highest effectiveness regarding MAP is almost always obtained in earlier fine-tuning steps.}

%The graph indicates that the fine-tuned model has shown saturated effectiveness after 100 fine-tuning epochs, shown as a non-statistically significant result than the best fine-tuning step. In addition, the result shows that theoretically, the fine-tuned neural models still have the potential to achieve even higher effectiveness, as the highest effectiveness regarding MAP is almost always obtained in earlier fine-tuning steps.

\subsection{Document Representation}

Document representation was title only (Title) and title and abstract (TiAb)\footnote{Note the results of the TiAb representation correspond to those also reported in Table~\ref{table:search_result_tiab}.} representations.
Table~\ref{table:search_result_title} shows how document representation impacts neural rankers' effectiveness. We only consider fine-tuned rankers, as zero-shot rankers did not correspond to viable effectiveness for the screening prioritisation task.

%, as described in section \ref{doc_rep}.
%
From these results, we find that using title and abstract within the neural rankers significantly outperforms using the title only representation, regardless of the underlying pre-trained language model employed. This finding generalises across all CLEF TAR datasets and all evaluation metrics, except for Recall@1\% for the BERT ranker on CLEF TAR 2017. This finding indicates that the abstracts contain essential information for the task. 

% find that using \textit{Tiab} as document representation in neural methods significantly outperforms using \textit{Title} representation. The finding generalises on all CLEF TAR datasets in all evaluation metrics, except in fine-tuned BERT model regarding Recall@1\% in CLEF TAR 2017. This finding indicates that abstracts contain essential information on the documents, and using \textit{Title} is not enough to represent candidate documents effectively. 

\begin{table*}[t]
	\centering
%	\small
	\begin{tabular}{ll|ll|llll|ll}
		\toprule
		\multicolumn{1}{l}{Dataset}&\multicolumn{1}{l|}{Method}&\multicolumn{1}{c}{Last\_Rel}&\multicolumn{1}{c|}{AP}&\multicolumn{1}{c}{Recall@1\%}&\multicolumn{1}{c}{Recall@5\%}&\multicolumn{1}{c}{Recall@10\%}&\multicolumn{1}{c|}{Recall@20\%}&\multicolumn{1}{c}{WSS95}&\multicolumn{1}{c}{WSS100} \\ \midrule
		
		\multirow{6}{*}{2017}&BERT-Tuned-TiAb&\textbf{2418.8333}&\textbf{0.2273}&0.1066&\textbf{0.4088}&\textbf{0.5888}&\textbf{0.7789}&\textbf{0.5435}&\textbf{0.4340}\\
		&BERT-Tuned-Title&3540.6667$^\dagger$&0.2037&\textbf{0.1119}&0.3429&0.4819$^\dagger$&0.6468$^\dagger$&0.2626$^\dagger$&0.1700$^\dagger$\\ \cmidrule{2-10}
		
		&BERT-M-Tuned-TiAb&\textbf{2475.2000}&\textbf{0.2770}&\textbf{0.1419}&\textbf{0.3727}&\textbf{0.5843}&\textbf{0.7707}&\textbf{0.5234}&\textbf{0.4187}\\
		&BERT-M-Tuned-Title&2953.4667&0.2299$^\dagger$&0.1082&0.3388&0.5084$^\dagger$&0.7027&0.4361$^\dagger$&0.3043$^\dagger$\\  \cmidrule{2-10}
		
		&BioBERT-Tuned-TiAb&\textbf{1461.6667}&\textbf{0.3078}&\textbf{0.1845}&\textbf{0.4903}&\textbf{0.6816}&\textbf{0.8355}&\textbf{0.6530}&\textbf{0.5913}\\
		&BioBERT-Tuned-Title&2072.1000$^\dagger$&0.2789$^\dagger$&0.1565&0.4176$^\dagger$&0.5937$^\dagger$&0.7919&0.5876$^\dagger$&0.4655$^\dagger$\\
		\midrule
		
		\multirow{6}{*}{2018}&BERT-Tuned-TiAb&\textbf{5581.7667}&\textbf{0.3467}&\textbf{0.1981}&\textbf{0.5028}&\textbf{0.6772}&\textbf{0.8196}&\textbf{0.6188}&\textbf{0.4121}\\
		&BERT-Tuned-Title&6087.9667&0.2652$^\dagger$&0.1303$^\dagger$&0.4054$^\dagger$&0.5667$^\dagger$&0.7328$^\dagger$&0.4898$^\dagger$&0.3004$^\dagger$\\
		\cmidrule{2-10}
		
		&BERT-M-Tuned-TiAb&\textbf{5185.5000}&\textbf{0.3387}&\textbf{0.1934}&\textbf{0.4833}&\textbf{0.6515}&\textbf{0.8265}&\textbf{0.6559}&\textbf{0.4815}\\
		&BERT-M-Tuned-Title&5757.7667$^\dagger$&0.2661$^\dagger$&0.1466$^\dagger$&0.3967$^\dagger$&0.5738$^\dagger$&0.7523$^\dagger$&0.5333$^\dagger$&0.3356$^\dagger$\\
		\cmidrule{2-10}
		
		&BioBERT-Tuned-TiAb&\textbf{4108.4000}&\textbf{0.4444}&\textbf{0.2768}&\textbf{0.5975}&\textbf{0.7574}&\textbf{0.8946}&\textbf{0.7194}&\textbf{0.6103}\\
		&BioBERT-Tuned-Title&4928.0000$^\dagger$&0.3557$^\dagger$&0.1862$^\dagger$&0.5202$^\dagger$&0.6952$^\dagger$&0.8524$^\dagger$&0.6527$^\dagger$&0.4813$^\dagger$\\
		\midrule

		\multirow{6}{*}{2019-dta}&BERT-Tuned-TiAb&\textbf{1399.3750}&\textbf{0.2234}&\textbf{0.1580}&\textbf{0.4390}&\textbf{0.6013}&\textbf{0.7620}&\textbf{0.5870}&\textbf{0.4600}\\
		&BERT-Tuned-Title&1597.2500&0.1851&0.1436&0.3802&0.4993$^\dagger$&0.6708$^\dagger$&0.5247&0.3892\\
		 \cmidrule{2-10}

		&BERT-M-Tuned-TiAb&\textbf{1178.0000}&\textbf{0.2535}&\textbf{0.2049}&\textbf{0.4474}&\textbf{0.5904}&\textbf{0.7536}&\textbf{0.6151}&\textbf{0.4997}\\
		&BERT-M-Tuned-Title&1798.6250&0.1858$^\dagger$&0.1195&0.3250$^\dagger$&0.5229&0.6831&0.5136&0.3729\\
		 \cmidrule{2-10}
		
		&BioBERT-Tuned-TiAb&\textbf{852.7500}&\textbf{0.3177}&\textbf{0.2604}&\textbf{0.4998}&\textbf{0.6710}&\textbf{0.8171}&\textbf{0.6857}&\textbf{0.5845}\\
		&BioBERT-Tuned-Title&1091.0000&0.2566$^\dagger$&0.1834&0.4761&0.6308&0.7876&0.6244&0.5095\\
		
		\midrule
		
		\multirow{6}{*}{2019-int.}&BERT-Tuned-TiAb&\textbf{1374.3000}&\textbf{0.2808}&\textbf{0.1646}&\textbf{0.3736}&\textbf{0.5274}&\textbf{0.6586}&\textbf{0.3629}&\textbf{0.3011}\\
		&BERT-Tuned-Title&1883.8500&0.2476&0.0953$^\dagger$&0.3231&0.4609&0.6134&0.3103&0.2763\\
		 \cmidrule{2-10}

		&BERT-M-Tuned-TiAb&\textbf{1571.2500}&\textbf{0.3343}&\textbf{0.1614}&\textbf{0.4021}&\textbf{0.5649}&\textbf{0.7061}&\textbf{0.4461}&\textbf{0.3623}\\
		&BERT-M-Tuned-Title&1653.6500&0.2702$^\dagger$&0.1045$^\dagger$&0.3488&0.5176&0.6891&0.3924&0.3288\\
		\cmidrule{2-10}
		
		&BioBERT-Tuned-TiAb&\textbf{706.8500}&\textbf{0.4559}&\textbf{0.2155}&\textbf{0.5805}&\textbf{0.7374}&\textbf{0.8417}&\textbf{0.6462}&\textbf{0.5794}\\
		&BioBERT-Tuned-Title&1145.5000$^\dagger$&0.3677$^\dagger$&0.1694&0.4868$^\dagger$&0.6447$^\dagger$&0.7572$^\dagger$&0.5222$^\dagger$&0.4622\\
		
		\bottomrule
		
	\end{tabular}
	\caption{Comparison of using \textit{Title} vs \textit{TiAb} as document representation. Statistical significance (Student's two-tailed paired t-test p < 0.05) between representation of two models is indicated by $\dagger$.}
	\label{table:search_result_title}
	\vspace{-20pt}
\end{table*}

\subsection{Topic by topic analysis}
\begin{figure*}[h]
	%	\begin{minipage}{\textwidth}
		\centering
		\includegraphics[width=0.65\linewidth]{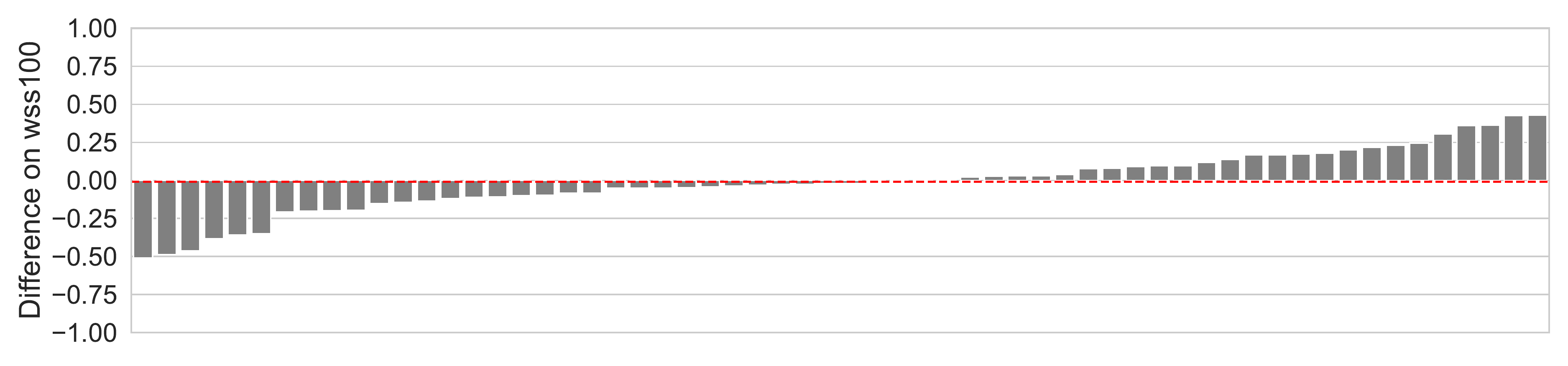}
%	\end{minipage}
	
%	\begin{minipage}{\textwidth}
		\centering
		\includegraphics[width=0.65\linewidth]{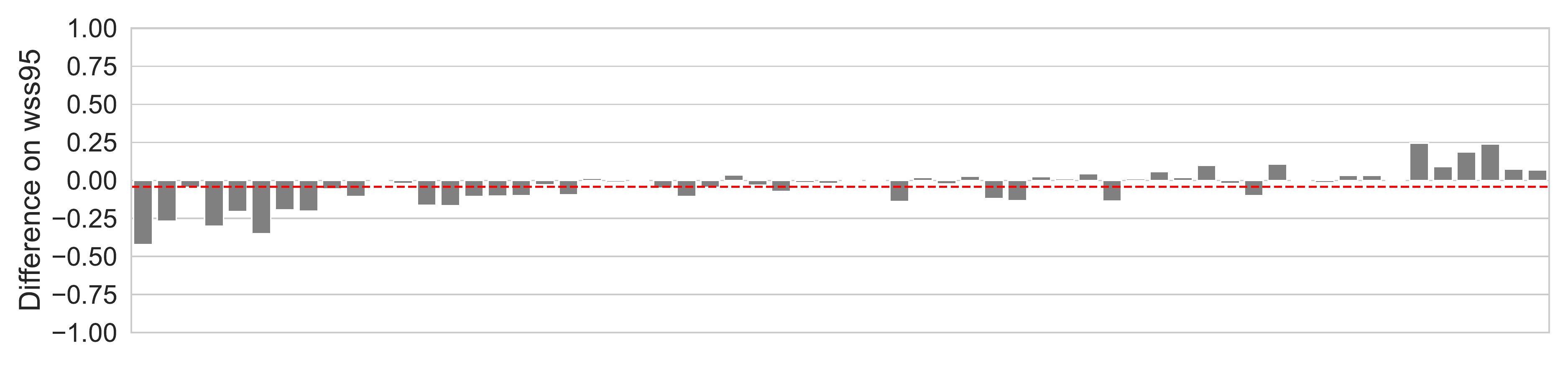}
%	\end{minipage}

%	\begin{minipage}{\textwidth}
	\centering
	\includegraphics[width=.65\linewidth]{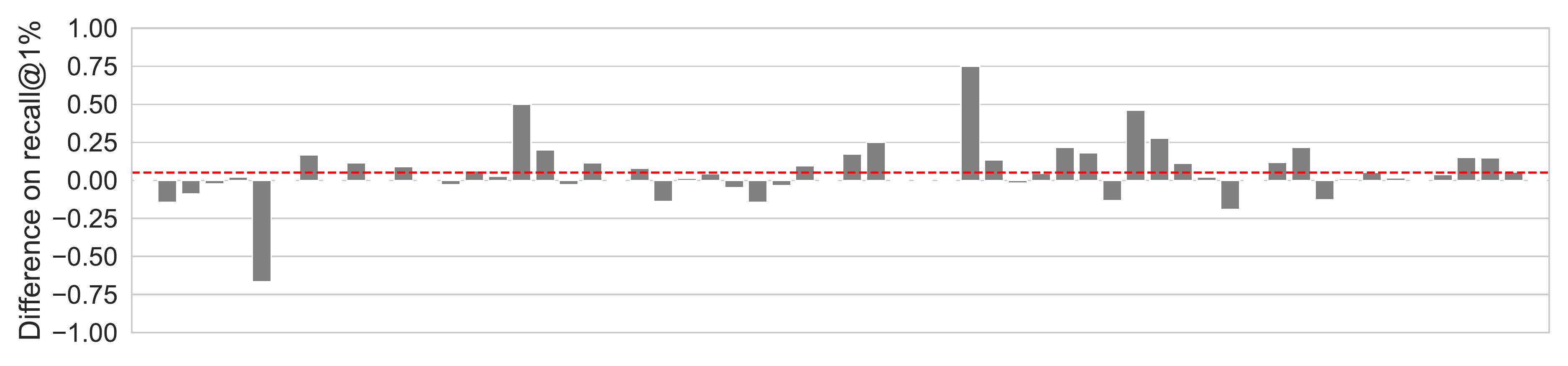}
%	\end{minipage}
	\caption{Per-topic effectiveness difference between BioBERT fine-tuned model and the best iterative runs for in CLEF TAR 2017 and 2018 (i.e., best active learning result of the year).}
	\label{fig:topic_analysis}
	
\end{figure*}

The results in Section~\ref{sec_fine_tuned} and Table~\ref{table:search_result_tiab} show that the fine-tuned BioBERT ranker achieves comparable effectiveness to the best iterative methods submitted to the CLEF tasks. We note this result is achieved by averaging effectiveness across topics in each dataset, and the datasets have very few topics (60 topics overall in 2017 and 2018). Thus, the average may be highly influenced by outliers, and because of this, we perform a deeper, topic-by-topic analysis of the results. In particular, we compare the BioBERT ranker to the best iterative run for the respective CLEF datasets. This analysis is shown in Figure \ref{fig:topic_analysis} as a gain-loss plot. 

%However, the performance comparison on the dataset is only based on 30 topics in each dataset (CLEF TAR 2017 and 2018); The average performance may be due to one or a few outlier cases. Therefore, we show topical comparison using a gain-loss plot in Figure \ref{fig:topic_analysis}. 

%\todo{@harry: this needs a better explanation (see slack).}
Topics with high effectiveness differ significantly between the two methods. Nearly half of the topics obtain higher effectiveness using the neural ranker that does not exploit feedback. 
Methods that use feedback will rely heavily on high effectiveness at early ranks. 
This is somewhat captured by comparing the WSS100 metric (a deep metric) with the  Recall@1\% metric (a relatively shallow metric, as it stops considering the ranking once 1\% of the relevant documents have been retrieved, which in most cases is very early into the ranking and quite far from the end of the ranking). Similarly, observations can be made if precision at rank 5 (P@5, another shallow metric) was considered (not shown in the figure). 

This illustrates that relevant feedback may not be needed for some topics as using neural rankers can already achieve significantly higher performance. Furthermore, this finding suggests that some topics may need feedback to ensure a higher screening prioritisation effectiveness. However, most relevance feedback ranking pipelines rely on the effectiveness of early ranked documents, especially in systematic review screening prioritisation, as relevant documents signals may be much harder to get than irrelevant ones. In Figure \ref{fig:topic_analysis}, we use Recall@1\% to show the effectiveness of early ranks; we find that on the topic level, neural rankers achieved higher performance in the majority of the topics, suggesting that neural rankers may give much better relevance signal in early ranks, thus lead to further improvements when the iterative method is applied.

%\subsection{Case study}

%\input{discussion}
\section{Conclusion}

In this paper, we investigate the effectiveness of rankers based on pre-trained language models for the task of screening prioritisation for systematic review creation. We focus on non-iterative rankers: those that produce a one-off ranking. In this context, we investigated neural rankers across two settings: zero-shot (apply the neural models without further training) and fine-tuned (apply the neural models after further training).

%using pre-trained language models directly to the task (zero-shot) or fine-tuning the model to the task (fine-tuned) to perform screening prioritisation. 

Our experiments show that while zero-shot neural rankers perform poorly, 
rankers fine-tuned on even a small amount of training data achieve significantly higher effectiveness than the current state-of-the-art non-iterative methods and comparable effectiveness to this task's current state-of-the-art iterative ranking methods. 
We also experimented with two different document representations (title only and title and abstract) and found that the abstract was essential for effective ranking.

%abstract encoding contains important information that leads to a significant effectiveness boost than only encoding the title of candidate documents. 

An interesting finding is that state-of-the-art iterative methods are better than the neural rankers for topics with good-quality initial rankings. 
On the other hand, the neural methods provide far better early rankings for most of the topics and, for many of these topics, these early wins result in overall better deep metrics (WSS, AP). It seems reasonable then to hypothesise that neural rankers could be further improved if cast into the iterative setting. For this to be possible, effective ways to exploit relevance assessments in the context of these neural rankers are required. We note that Yang et al~\cite{yang2022goldilocks} have proposed a method for iteratively exploiting relevance assessments in the context of a  classifier based on pre-trained language models for Technology Assisted Review. This classifier bears similarity with our neural rankers. However, a clear drawback of their method is the high computational costs involved with their continuous learning approach and the consequent high latency imposed on the user. We highlight that expert screeners can return an assessment every 20--30 seconds~\cite{clark2020full, wallace2010semi}, a timeframe sensibly lower than the latency of Yang et al.'s method (1-2 hours). An alternative direction is considering methods for relevance feedback in the context of rankers based on pre-trained language models. Two main classes of methods have been proposed in this respect. The first class of methods combines the text input of the feedback with that of the query and the document to be scored~\cite{padaki2020rethinking, wang2020pseudo, wang2022neural}: these approaches are severely impacted by the language model's input size limit and would not be viable for the screening prioritisation task. The second class of methods instead combines the representations of feedback documents, not their text~\cite{hang2022relevance, hang2021improving, yu2021improving, li2021pseudo, wang2021pseudo}: these methods trade-off some loss in effectiveness compared to the first class of methods for the ability to model indefinitely long feedback and for lower latency and computational costs. Immediate future work on the use of rankers based on pre-trained language models for screening prioritisation should be directed towards investigating and adapting these two classes of approaches.

Neural rankers show promise in terms of effectiveness for screening prioritisation in systematic review creation. These rankers have the potential to greatly reduce the effort of compiling systemic reviews. This impact is better quality and a greater number of systematic reviews being produced, which improves the important medical and policy decisions based on these reviews.

\subsection*{Acknowledgements.}  Shuai Wang is supported by a UQ Earmarked PhD Scholarship and this research is funded by the Australian Research Council Discovery Project DP210104043.

\balance

\bibliographystyle{ACM-Reference-Format}
\bibliography{references.bib}

\end{document}